\documentclass[10pt]{elsarticle}
\usepackage{amssymb}
\usepackage{amsmath}
\usepackage{graphicx}
\usepackage{dcolumn}
\usepackage{bm}
\newcommand{\rar}{\rightarrow}
\newcommand{\ra}{\rangle}
\newcommand{\la}{\langle}
\newcommand{\ttt}{\texttt}
\newcommand{\mb}{\mathbb}

\journal{Knowledge-Based Systems}

\begin{document}

\begin{frontmatter}

\title{Grammar-Based Geodesics in Semantic Networks\footnote{Rodriguez, M.A., Watkins, J., ``Grammar-Based Geodesics in Semantic Networks," Knowledge-Based Systems, 23(8), pp. 844--855, doi:10.1016/j.knosys.2010.05.009, December 2010.}}

\author{Marko A. Rodriguez}
\address{T-7, Center for Nonlinear Studies \\
		Los Alamos National Laboratory \\
		Los Alamos, New Mexico 87545 }

\author{Jennifer H. Watkins}
\address{International and Applied Technology \\
		Los Alamos National Laboratory \\
		Los Alamos, New Mexico 87545 }
		
\date{\today}
\begin{abstract}
A geodesic is the shortest path between two vertices in a connected network. The geodesic is the kernel of various network metrics including radius, diameter, eccentricity, closeness, and betweenness. These metrics are the foundation of much network research and thus, have been studied extensively in the domain of single-relational networks (both in their directed and undirected forms). However, geodesics for single-relational networks do not translate directly to multi-relational, or semantic networks, where vertices are connected to one another by any number of edge labels. Here, a more sophisticated method for calculating a geodesic is necessary. This article presents a technique for calculating geodesics in semantic networks with a focus on semantic networks represented according to the Resource Description Framework (RDF). In this framework, a discrete ``walker" utilizes an abstract path description called a grammar to determine which paths to include in its geodesic calculation. The grammar-based model forms a general framework for studying geodesic metrics in semantic networks.
\end{abstract}

\end{frontmatter}

\section{Introduction}

The study of networks (i.e. graph theory) is the study of the relationship between vertices (i.e.~nodes) as defined by the edges (i.e.~arcs) connecting them. In path analysis, a path metric function maps an ordered vertex pair into a real number, where that real number is the length of the path connecting to the two vertices. Metrics that utilize the shortest path between two vertices in their calculation are called geodesic metrics. The geodesic metrics that will be reviewed in this article are shortest path, eccentricity \cite{eccen:harary1995}, radius, diameter, betweenness centrality \cite{between:freeman1977}, and closeness centrality \cite{close:bavelas1950}. 

If $G^1$ is a single-relational network, then $G^1 = (V, E)$, where $V = \{i,\ldots,j \},$ is the set of vertices and $E \subseteq (V \times V)$ is a subset of the product of $V$. In a single-relational network all the edges have a single, homogenous meaning. Because an edge in a single-relational network is an element of the product of $V$, it does not have the ability to represent the type of relationships that exist between the two vertices it connects. An edge can only denote that there is a relationship. Without a distinguishing label, all edges in such networks have a single meaning. Thus, they are called single-relational networks.\footnote{It is noted that bipartite networks allow for more than one edge meaning to be inferred because $V$ is the union of two disjoint vertex sets. Thus, edges from set $A \subset V$ to set $B \subset V$ (such that $A \cap B = \emptyset$) can have a different meaning than the edges from $B$ to $A$. Also, theoretically, it is possible to represent edge labels as a topological feature of the graph structure \cite{mapnetwork:rodriguez2009}. In other words, there exists an injective function (though not surjective) from the set of semantic networks to the set of single-relational networks that preserves the meaning of the edge labels.} While a single-relational network supports the representation of a homogeneous set of relationships, a semantic network supports the representation of a heterogeneous set of relationships. For instance, in a single-relational network it is possible to represent humans connected to one another by friendship edges; in a semantic network, it is possible to represent humans connected to one another by friendship, kinship, collaboration, communication, etc. relationships. 

A semantic network denoted $G^n$ can be defined as a set of single-relational networks such that $G^{n} = (V, \mb{E})$, where $\mb{E} = \{E_0, E_1, \ldots, E_n \}$ and for any $E_k \in \mb{E}$, $E_k \subseteq (V \times V)$ \cite{netanal:brandes2005}. The meaning of a relationship in $G^n$ is determined by its set $E_k \in \mb{E}$. Perhaps a more convenient semantic network representation and the one to be used throughout the remainder of this article is that of the triple list where $G^{n} \subseteq (V \times \Omega \times V)$ and $\Omega$ is a set of edge labels. A single edge in this representation is denoted by a triple $\tau = \la i, \omega, j \ra$, where vertex $i$ is connected to vertex $j$ by the edge label $\omega$.

In some cases, it is possible to isolate sub-networks of a semantic network and represent the isolated network in an unlabeled form. Unlabeled geodesic metrics can be used to compute on the isolated component. However, in many cases,  the complexity of the path description does not support an unlabeled representation. These scenarios require ``semantically aware" geodesic metrics that respect a semantic network's ontology (i.e.~the vertex classes and edge types) \cite{socialgrammar:rodriguez2007}. A semantic network is not simply a directed labeled network. It is a high-level representation of complex objects and their relationship to one another according to ontological constraints. There exist various algorithms to study semantically typed paths in a network \cite{rhoquery:anyanwu2003,ranksem:zhuge2003,discov:lin2004,semrank:boan2005,semassoc:sheth2005}. Such algorithms assume only a path between two vertices and do not investigate other features of the intervening vertices. The benefit of the grammar-based geodesic model presented in this article is that complex paths can be represented to make use of path ``bookkeeping." Such bookkeeping investigates intervening vertices even though they may not be included in the final path solution. For example, it may be important to determine a set of ``friendship" paths between two human vertices, where every intervening human works for a particular organization and has a particular position in that organization. While a set of friendship paths is the result of the function, the path detours to determine employer and position are not. The technique for doing this is the primary contribution of this article. 

A secondary contribution is the unification of the grammar-based model proposed here with the grammar-based model proposed in \cite{grammar:rodriguez2008} for calculating stationary probability distributions in a subset of the full semantic network (e.g.~eigenvector centrality  \cite{power:bonacich1987} and PageRank \cite{anatom:brin1998}). With the grammar-based model, a single framework exists that ports many of the popular single-relational network analysis algorithms to the semantic network domain. Moreover, an algebra for mapping semantic networks to single-relational networks has been presented in \cite{pathalg:rodriguez2009} and can be used to meaningfully execute standard single-relational network analysis algorithms on distortions of the original semantic network. The Semantic Web community does not often employee the standard suite of network analysis algorithms. This is perhaps due to the fact that the Semantic Web is generally seen as a knowledge-base grounded in description logics rather than graph- or network-theory. When the Semantic Web community adopts a network interpretation, it can benefit from the extensive body of work found in the network analysis literature. For example, recommendation \cite{kbsrec:blanco2008}, ranking \cite{chitra:nodelabel2004}, and decision making \cite{socialgrammar:rodriguez2007} are a few of the types of Semantic Web applications that can benefit from a network perspective. In other words, graph/network theoretic techniques can be used to yield innovative solutions on the Semantic Web.

The first half of this article will define a popular set of geodesic metrics for single-relational networks. It will become apparent from these definitions, that the more advanced geodesics rely on the shortest path metric. The second half of the article will present the grammar-based model for calculating a meaningful shortest path in a semantic network. The other geodesics follow from this definition.

\section{Geodesics in Single-Relational Networks\label{sec:geos}}

This section will review a collection of popular geodesic metrics used to characterize a path, a vertex, and a network. The following list enumerates these metrics and identifies whether they are path, vertex, or network metrics:
\begin{itemize}\addtolength{\itemsep}{-0.7\baselineskip}
	\item in- and out-degree: vertex metric
	\item shortest path: path metric
	\item eccentricity: vertex metric
	\item radius: network metric
	\item diameter: network metric
	\item closeness: vertex metric
	\item betweenness: vertex metric.
\end{itemize}

It is worth noting that besides in- and out-degree, all the metrics mentioned utilize a path function $\rho: V \times V \rar Q$ to determine the set of paths between any two vertices in $V$, where $Q$ is a set of paths. The premise of this article is that once a path function is defined for a semantic network, then all of the other metrics are directly derived from it. In the semantic network path function, $\rho: V \times V \times \Psi \rar Q$ returns the number of paths between two vertices according to a user-defined grammar $\Psi$.

Before discussing the grammar-based geodesic model for semantic networks, this section will review the geodesic metrics in the domain of single-relational networks.

\subsection{In- and Out-Degree}

The simplest structural metric for a vertex is the vertex's degree. While this is not a geodesic metric, it is presented as the concept will become necessary in the later section regarding semantic networks. 

For directed networks, any vertex $i \in V$ has both an in-degree and an out-degree. The set of edges in $E$ that have $i$ as either its in- or out-edge is denoted $\Gamma^-: V \rar E$ and $\Gamma^+: V \rar E$, respectively. If 
\begin{equation*}
	\Gamma^-(i) = \{ (x,y) \; | \; (x,y) \in E \; \wedge \; y = i \}
\end{equation*}
and 
\begin{equation*}
	\Gamma^+(i) = \{ (x,y) \; | \; (x,y) \in E \; \wedge \; x = i \}
\end{equation*}
 then, $\Gamma^-(i)$ is the subset of edges in $E$ incoming to $i$ and  $\Gamma^+(i)$ is the subset of edges outgoing from $i$. The cardinality of the sets is the in- and out-degree of the vertex, denoted $|\Gamma^-(i)|$ and $|\Gamma^+(i)|$, respectively.

\subsection{Shortest Path}

The shortest path metric is the foundation for all other geodesic metrics. This metric is defined for any two vertices $i,j \in V$ such that the sink vertex $j$ is reachable from the source vertex $i$ in $G^1$ \cite{short:dijkstra1959}. If $j$ is unreachable from $i$, the shortest path between $i$ and $j$ is undefined. The shortest path between any two vertices $i$ and $j$ in an unweighted network is the smallest of the set of all paths between $i$ and $j$. If $\rho : V \times V \rar Q$ is a function that takes two vertices and returns a set of paths $Q$ where for any $q \in Q$, $q = (i,\dots,j)$, then the shortest path between $i$ and $j$ is the $min(\bigcup_{q \in Q} |q|-1)$, where $min$ returns the smallest value of its domain. The shortest path function is denoted $s : V \times V \rar \mb{N}$ with the function rule
\begin{equation*}
	s(i,j) = min\left(\bigcup_{q \in \rho(i,j)} |q|-1\right).
\end{equation*}
It is important to subtract $1$ from the path length since a path is defined as the set of edges traversed, not the set of vertices traversed. Thus, for the path $q = (a,b,c,d)$, the $|q|$ is $4$, but the path length is $3$.

Note that $\rho$ returns the set of all paths between $i$ and $j$. Of course, with the potential for loops, this function could return a $|Q| = \infty$. Therefore, in many cases, it is important to not consider all paths, but just those paths that have the same cardinality as the shortest path currently found and thus are shortest paths themselves. It is noted that all the remaining geodesic metrics require only the shortest path between $i$ and $j$.

\subsection{Eccentricity, Radius, and Diameter}

The radius and diameter of a network require the determination of the eccentricity of every vertex in $V$. The eccentricity metric requires the calculation of $|V| - 1$ shortest path calculations of a particular vertex \cite{eccen:harary1995}. The eccentricity of a vertex $i$ is the largest shortest path between $i$ and all other vertices in $V$  such that the eccentricity function $e : V \rar \mb{N}$ has the rule
\begin{equation*}
	e(i) = max\left(\bigcup_{j \in V} s(i,j)\right),
\end{equation*}
where $max$ returns the largest value of its domain.

The radius of the network is the minimum eccentricity of all vertices in $V$ \cite{socialanal:wasserman1994}. The function $r : G \rar \mb{N}$ has the rule 
\begin{equation*}
	r(G^1) = min\left(\bigcup_{i \in V} e(i)\right).
\end{equation*}

Finally, the diameter of a network is the maximum eccentricity of the vertices in $V$ \cite{socialanal:wasserman1994}. The function $d : G \rar \mb{N}$ has the rule 
\begin{equation*}
	d(G^1) = max\left(\bigcup_{i \in V} e(i)\right).
\end{equation*}

\subsection{Closeness and Betweenness Centrality}

Closeness and betweenness centrality are popular network metrics for determining the ``centralness" of a vertex. Closeness centrality is defined as the mean shortest path between some vertex $i$ and all the other vertices in $V$  \cite{close:bavelas1950,close:leavitt1951,close:sabaidussi1966}. The function $c : V \rar \mb{R}$ denotes the closeness function and has the rule
\begin{equation*}
	c(i) = \frac{1}{\sum_{j \in V} s(i,j)}.
\end{equation*}

Betweenness centrality is defined for a vertex in $V$. The betweenness of $i \in V$ is the number of shortest paths that exist between all vertices $j \in V$ and $k \in V$ that have $i$ in their path divided by the total number of shortest paths between $j$ and $k$, where $i \neq j \neq k$ \cite{between:freeman1977,betweeness:brandes2001}. If $\sigma : V \times V \rar Q$ is a function that returns the set of shortest paths between any two vertices $j$ and $k$ such that 
\begin{equation*}
	\sigma(j,k) = \bigcup_{q \in p(j,k)} q : |q|-1 = s(j,k)
\end{equation*}
and $\hat{\sigma} : V \times V \times V \rar Q$ is the set of shortest paths between two vertices $j$ and $k$ that have $i$ in the path, where
\begin{equation*}
	\hat{\sigma}(j,k,i) = \bigcup_{q \in p(j,k)} q : (|q|-1 = s(j,k) \; \wedge \; i \in q), 
\end{equation*}
then the betweenness function $b : V \rar \mb{R}$ has the rule
\begin{equation*}
	b(i) = \sum_{i \neq j \neq k \in V} \frac{|\hat{\sigma}(j,k,i)|}{|\sigma(j,k)|}
\end{equation*}

It is worth noting that in \cite{between:newman2003}, the author articulates the point that the shortest paths between two vertices is not necessarily the only mechanism of interaction between two vertices. Thus, the author develops a variation of the betweenness metric that favors shortest paths, but does not utilize only shortest paths in its betweenness calculation.

\section{Semantic Network Grammars}

A semantic network is a directed labeled graph. However, a semantic network is perhaps best interpreted in an object-oriented fashion where complex objects (i.e.~multi-vertex elements) are connected to one another according to various relationship types. While a particular human is represented by a vertex, metadata associated with that individual is represented in the vertices adjacent to the human vertex (e.g.~the human's name, address, age, etc.). In many instances, particular metadata vertices are sinks (i.e.~no outgoing edges). In other cases, the metadata of an individual is another complex object such as the friend of that human or the human's employer.

The topological features of a semantic network are represented by a data type abstraction called an ontology (i.e.~a semantic network schema). A popular semantic network representation is the Resource Description Framework (RDF) \cite{rdfspec:manola2004}. RDF Schema (RDFS) is a schema language for developing RDF ontologies in RDF \cite{rdfs:brickley2004}. This article will present all of its concepts from the perspective of RDF and RDFS primarily due to the fact that these are standard data models with a large application-base. However, these ideas can be generalized to any semantic network representation. This is due to the fact that one can remove the constraint of using URIs, literals, and blank nodes when labeling vertices and edges. When such a constraint is lifted, then a directed, vertex/edge-labeled, multi-graph results. In the semantic network literature, such an abstract graph type is named a semantic network \cite{semdef:sowa1987}. The first subsection will briefly introduce the concept of RDF and RDFS before describing an ontology for designing geodesic grammars.

\subsection{Introduction to RDF/RDFS}

The RDF data model represents a semantic network as a triple list where the vertices and edges (both called resources) are Uniform Resource Identifiers (URI) \cite{uri:berners2005}, blank nodes, or literals. If the set of all URIs is denoted $U$, the set of all blank nodes is denoted $B$, and the set of all literals is denoted $L$, then an RDF network is the triple list $G^n$ such that
\begin{equation*}
	G^n \subseteq ((U \cup B) \times U \times (U \cup B \cup L)).
\end{equation*}
The first resource of a triple is called the subject, the second is called the predicate, and the third is called the object. A single triple $\tau \in G^n$ is denoted as $\tau = \la s, p, o \ra$. 

All URIs are namespaced such that the URI \ttt{http://www.lanl.gov\#marko} has a namespace of \ttt{http://www.lanl.gov\#} and a fragment of \ttt{marko}. In many cases, for document and diagram clarity, a namespace is prefixed in such a way that the previous URI is represented as \ttt{lanl:marko}. In this article, the namespaces for RDF and RDFS will be prefixed as \ttt{rdf} and \ttt{rdfs}, respectively. 

Blank nodes are ``anonymous" vertices and are not discussed in this article as they will not directly pertain to any of the concepts presented. Literals are any resource that denotes a string, integer, floating point, date, etc. The full taxonomy of literal types is presented in \cite{xsd:biron2004}.

In RDFS, every vertex is tied to some platonic category representing its \ttt{rdfs:Class} using the \ttt{rdf:type} property. Moreover, every edge label has domain/range restrictions that determine the vertex types that the edge labels can be used in conjunction with. Because the instance of an ontology obeys the defined constraints of the ontology, the modeler has an abstract representation of the topological features of the semantic network instance in terms of classes (vertices) and properties (edge labels). For example,
\begin{align*}
& \la \ttt{lanl:hasFriend}, \ttt{rdfs:domain}, \ttt{lanl:Human} \ra \\
& \la \ttt{lanl:hasFriend}, \ttt{rdfs:range}, \ttt{lanl:Human} \ra
\end{align*}
states that any resource of type \ttt{lanl:Human} can have a friend that is only of type \ttt{lanl:Human}. Therefore, the following three triples are legal according to the simple ontology above:
\begin{align*}
& \la \ttt{lanl:marko}, \ttt{rdf:type}, \ttt{lanl:Human} \ra \\
& \la \ttt{lanl:jen}, \ttt{rdf:type}, \ttt{lanl:Human} \ra \\
& \la \ttt{lanl:marko}, \ttt{lanl:hasFriend}, \ttt{lanl:jen} \ra .
\end{align*}
However, the three statements
\begin{align*}
& \la \ttt{lanl:marko}, \ttt{rdf:type}, \ttt{lanl:Human} \ra \\
& \la \ttt{lanl:fluffy}, \ttt{rdf:type}, \ttt{lanl:Dog} \ra \\
& \la \ttt{lanl:marko}, \ttt{lanl:hasFriend}, \ttt{lanl:fluffy} \ra
\end{align*}
are not legal according to the ontology because \ttt{lanl:fluffy} is a \ttt{lanl:Dog} and a \ttt{lanl:Human} cannot befriend anything that is not a \ttt{lanl:Human}.

\begin{figure}[h!]
	\begin{center}
		\includegraphics[width=0.5\textwidth]{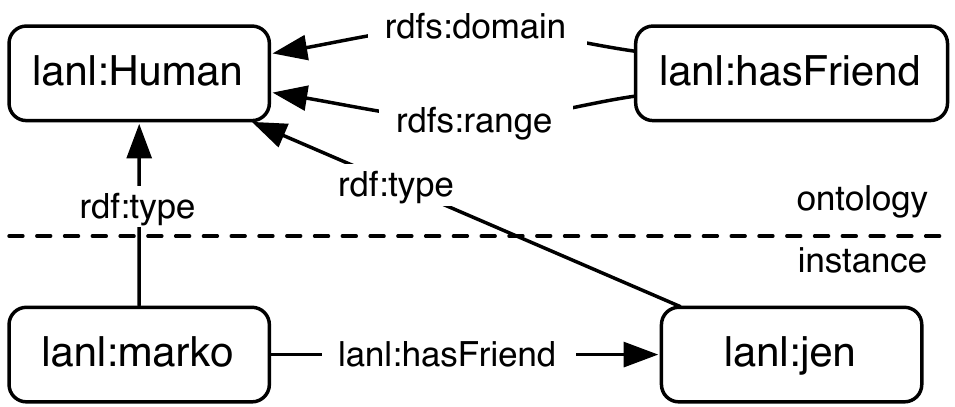}
	\caption{\label{fig:friend-full}The full representation of all triples in the ontology and instance layers of the semantic network example.}
	\end{center}
\end{figure}

\begin{figure}[h!]
	\begin{center}
		\includegraphics[width=0.5\textwidth]{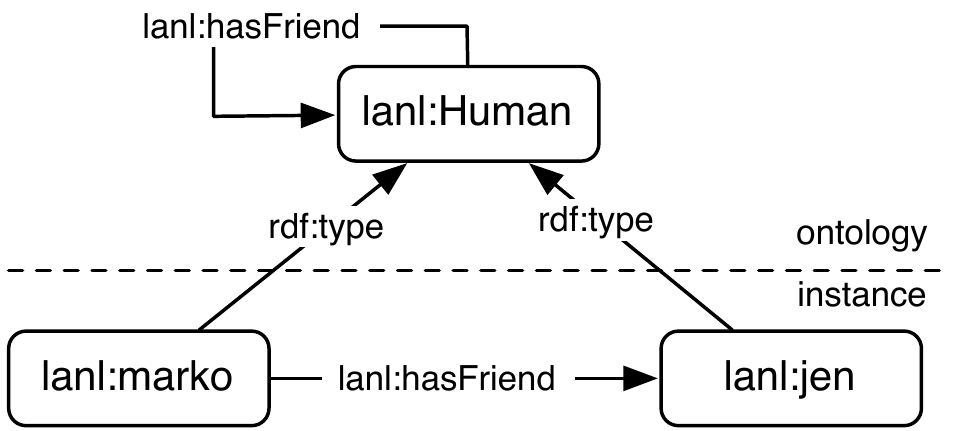}
	\caption{\label{fig:friend-abbrev}The abbreviated representation of the ontology and instance layers of the semantic network example.}
	\end{center}
\end{figure}

The ontology and legal instance of the previous example are diagrammed in Figure \ref{fig:friend-full}. However, for the sake of brevity and clarity of the diagram, the domain and range properties of a class can be abbreviated as in Figure \ref{fig:friend-abbrev}. The abbreviated ontological diagram will be used throughout the remainder of this article. It is important to note that both the RDFS ontology and RDF instance network are represented in RDF and thus, both instances and ontology are contained within a single semantic network.

Finally, an important concept in RDFS is \ttt{rdfs:Class} and \ttt{rdf:Property} subsumption as denoted by the \ttt{rdfs:subClassOf} and \ttt{rdfs:subPropertyOf} predicates, respectively. With the \ttt{rdfs:subClassOf} and \ttt{rdfs:subPropertyOf} predicates, it is possible to generate concept hierarchies. For the purposes of this article, it is only necessary to understand that subsumption is transitive such that if
\begin{align*}
& \la \ttt{lanl:fluffy}, \ttt{rdf:type}, \ttt{lanl:Dog} \ra \\
& \la \ttt{lanl:Dog}, \ttt{rdfs:subClassOf}, \ttt{lanl:Mammal} \ra \\
& \la \ttt{lanl:Mammal}, \ttt{rdfs:subClassOf}, \ttt{lanl:Animal} \ra,
\end{align*}
then it can be inferred that because \ttt{lanl:fluffy} is a \ttt{lanl:Dog}, \ttt{lanl:fluffy} is also both a \ttt{lanl:Mammal} and a \ttt{lanl:Animal}. Transitivity exists for the \ttt{rdfs:subPropertyOf} predicate as well.

\subsection{Defining a Grammar}

This subsection will define the RDFS ontology for creating a grammar. Any user-defined grammar must obey this ontology. The grammar constructed from this ontology determines the meaning of the value returned by a ``semantically aware" geodesic function. Any grammar instance is denoted $\Psi \subseteq ((U \times B) \times U \times (U \times B \times L))$.  

\begin{figure}[h!]
	\begin{center}
		\includegraphics[width=0.8\textwidth]{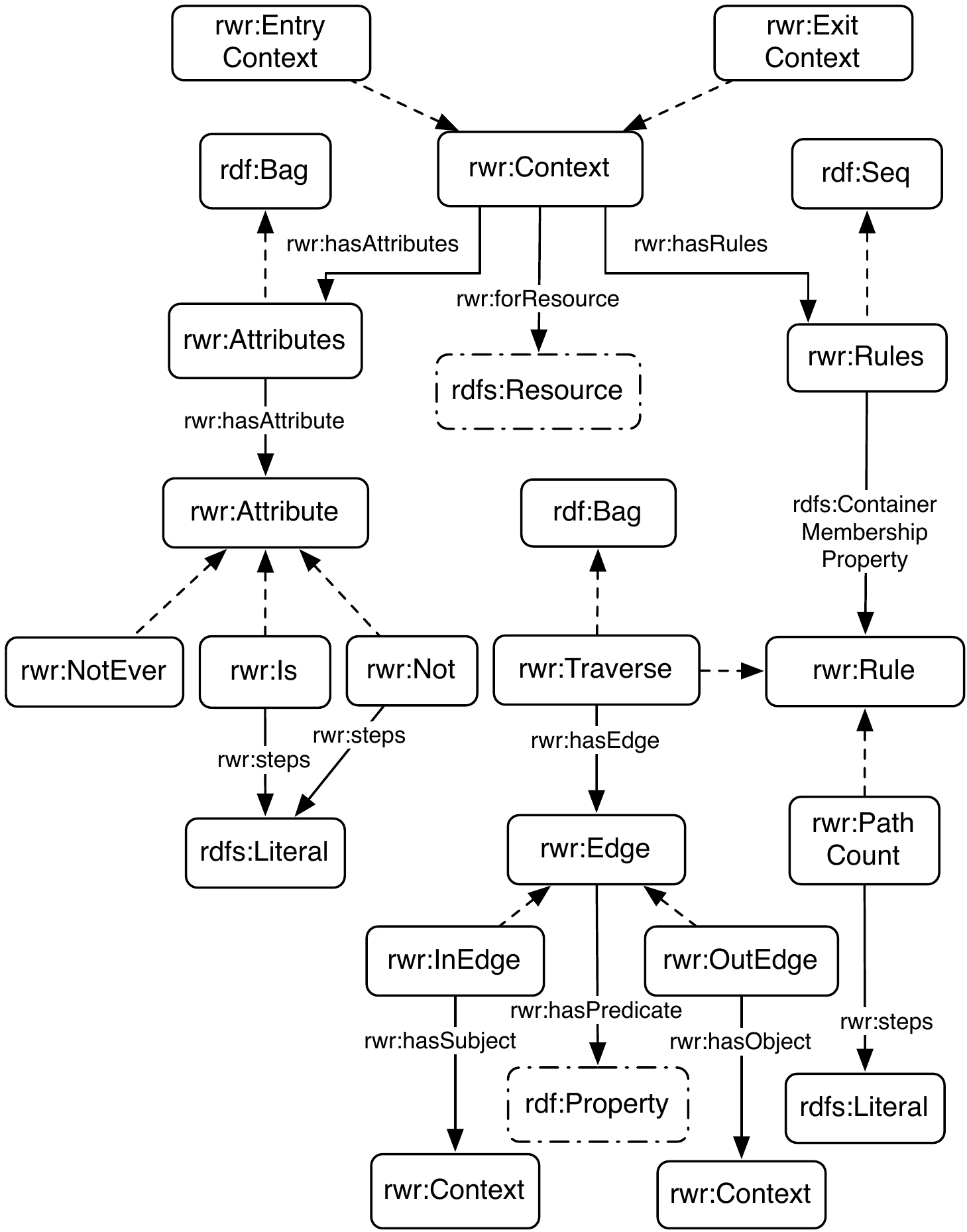}
	\caption{\label{fig:ont-rwr}The ontology for a geodesic path grammar.}
	\end{center}
\end{figure}

The instance of a grammar is represented in RDF and the ontology of the grammar is represented in RDFS. Figure \ref{fig:ont-rwr} diagrams the ontology of the geodesic grammar, where edges represent properties whose tail is the domain of the property and whose head is the range of the property. Furthermore, the dashed edges denote the RDFS property \ttt{rdfs:subClassOf}.

The remainder of this section will  present an informal review of the major components of the grammar ontology. The next section will formalize all aspects of the resources diagrammed in Figure \ref{fig:ont-rwr}.

Grammar-based geodesics rely on a discrete walker. The walker utilizes a $\Psi$ grammar to constrain its path through $G^n$. The combination of a walker and a $\Psi$ is a breadth-first search through a particular sub-network of $G^n$. That sub-network is abstractly represented by $\Psi$, but not fully realized until after the execution of $\Psi$ on $G^n$.

Any $\Psi$ is a collection of \ttt{rwr:Context} resources connected to one another by \ttt{rwr:Traverse} resources. Each \ttt{rwr:Context} is an abstract representation of a legal step along a path that a walker can traverse on its way from source vertex $i$ to sink vertex $j$. An \ttt{rwr:Context} has an associated \ttt{rwr:forResource} property. The object of that property determines the set of legal vertices that that the \ttt{rwr:Context} can resolve to. Only when a walker utilizes a grammar do the \ttt{rwr:Context}s have a resolution to a particular vertex in $G^n$. \ttt{rwr:Context} resolution is further constrained by the \ttt{rwr:Rules} and \ttt{rwr:Attributes} of the \ttt{rwr:Context} in $\Psi$. 

Two important data structures that are used in a grammar are the \ttt{rdf:Bag} and \ttt{rdf:Seq}. An \ttt{rdf:Bag} is an unordered set of elements where each element of the \ttt{rdf:Bag} is the object of a triple with predicate \ttt{rdf:li}. An \ttt{rdf:Seq} is an ordered set of elements where each element of the \ttt{rdf:Seq} is the object of a triple with a predicate that is an \ttt{rdfs:subPropertyOf} \ttt{rdfs:ContainerMembershipProperty} (i.e.~\ttt{rdf:\_1}, \ttt{rdf:\_2}, \ttt{rdf:\_3}, etc.).

There exist two \ttt{rwr:Rules} (an \ttt{rdfs:subClassOf} \ttt{rdf:Seq}): \ttt{rwr:PathCount} and \ttt{rwr:Traverse}. The \ttt{rwr:PathCount} rule instructs the walker to record the vertex, edge, and directionality in the ordered path set that is ultimately returned by the grammar-based geodesic algorithm. The \ttt{rwr:Traverse} rule instructs the walker to select some outgoing or incoming edge of its current vertex as defined by the set of \ttt{rwr:Edge}s associated with the \ttt{rwr:Traverse} rule. If more than one choice should exist for the walker, the walker chooses both by cloning itself and having each clone take a unique branch of the path.

There exist three \ttt{rwr:Attributes} (an \ttt{rdfs:subClassOf} \ttt{rdf:Bag}): \ttt{rwr:NotEver}, \ttt{rwr:Is}, and \ttt{rwr:Not}. In some instances, when traversing to a new vertex, the walker must respect the fact that it has already seen a particular vertex. The \ttt{rwr:NotEver} attribute ensures that the resolution of the \ttt{rwr:Context} is not a previously seen vertex, thus preventing infinite loops. The \ttt{rwr:Is} attribute allows the walker to explore an area around a particular vertex (i.e.~other paths not directly associated with the return path) while still ensuring that the walker returns to the original vertex. Finally, the \ttt{rwr:Not} attribute ensures that the walker does not return to a \textit{particular} previously seen vertex. 

\begin{figure}[h!]
	\begin{center}
		\includegraphics[width=0.6\textwidth]{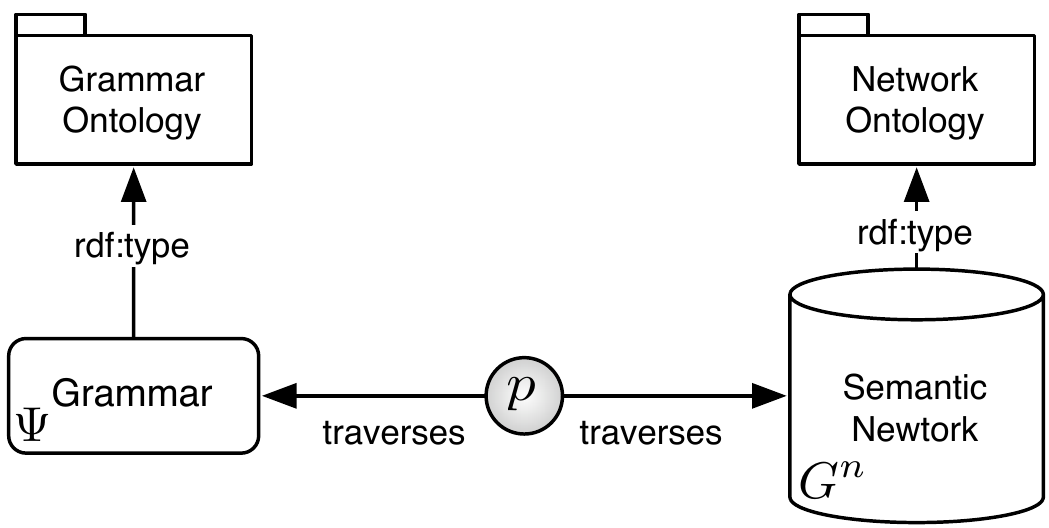}
	\caption{\label{fig:system} A walker $p$ walks both $\Psi$ and $G^n$.}
	\end{center}
\end{figure}

If vertex $i$ is the head of the path (i.e.~source), then it is defined in an \ttt{rwr:EntryContext}. If vertex $j$ is the tail of the path (i.e.~sink), then it is defined in an \ttt{rwr:ExitContext}. The purpose of the walker is to move from source to sink in $G^n$ by respecting the \ttt{rwr:Rules} and \ttt{rwr:Attributes} of the \ttt{rwr:Context}s that it traverses in $\Psi$. Figure \ref{fig:system} diagrams the relationship between a walker, its grammar $\Psi$, and its network instance $G^n$. The grammar acts as a user-defined ``program" that the walker executes, where the language of that program is defined by the grammar ontology.

The next section will formalize the grammar.

\section{Formalizing the Grammar-Based Model}

Once a grammar has been defined according to the constraints of the ontology diagrammed in Figure \ref{fig:ont-rwr}, the path function $\rho: V \times V \times \Psi \rar Q$ can be executed. The function $\rho$ returns the set of all paths between any two vertices $i,j \in V$. This section will define the rules by which $\rho$ interprets its domain parameters and ultimately derives a path set. 

The grammar-based model requires the walker to query $G^n$ such that it can determine the set of legal vertices and edges that it can traverse. Moreover, the walker must be able to query $\Psi$ in order to know which \ttt{rwr:Rule}s and \ttt{rwr:Attributes} to respect. The mechanism by which the walker queries $G^n$ and $\Psi$ is called the symbol binding model. For example, the following query
\begin{align*}
	X = & \{?x \; | \; \la ?x, \ttt{lanl:hasFriend}, \ttt{lanl:jhw} \ra \in G^n \\
	       & \; \wedge \la ?x, \ttt{lanl:worksFor}, \ttt{lanl:LANL} \ra \in G^n \} 
\end{align*}
would fill the unordered set $X$ with all people that have \ttt{lanl:jhw} as their friend and who work for \ttt{lanl:LANL}. A more advanced query example is
\begin{align*}
	X = & \{?x,?y \; | \; \la ?x, \ttt{lanl:hasFriend}, ?y \ra \in G^n \\
	       & \; \wedge \la ?y, \ttt{lanl:worksFor}, \ttt{lanl:LANL} \ra \in G^n \\
	       & \; \wedge \la ?x, \ttt{lanl:worksFor}, \ttt{lanl:PNNL} \ra \in G^n \} .
\end{align*}
In the above query, the set $X$ is an unordered set of ordered pairs of friends where one of the friends works at \ttt{lanl:LANL} and the other works at \ttt{lanl:PNNL}.

\subsection{Initializing a Walker $p$}

The path function $\rho$ is supplied with a start vertex $i$, an end vertex $j$, and a grammar $\Psi$. Upon the execution of $\rho$, a single walker, denoted $p$, is created and added to the set of walkers $P$, where at $n=0$, $|P|=1$, and $n \in \mb{N}$ is in discrete time. The set $P$ may increase in size over the course of the algorithm as clone particles are created where multiple legal options exist for traversal.

\begin{figure}[h!]
	\begin{center}
		\includegraphics[width=0.85\textwidth]{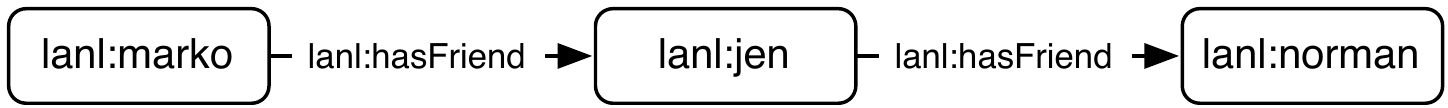}
	\caption{\label{fig:path-example}An example of a $g^p$ path.}
	\end{center}
\end{figure}

Every walker has two ordered multi-sets associated with it: $g^p$ and $q^p$. The multi-set $g^p$ is an ordered set of vertices, edges, and edge directions traversed by $p$, where $g^p_n$ is the vertex location of $p$ at time step $n$. The element $g^p_{n'}$ denotes the predicate (i.e.~edge label) used by $p$ to traverse to $g^p_n$ and the element $g^p_{n''}$ denotes the directionality of the predicate used in that traversal. For example, suppose $g^p = ($\ttt{lanl:marko}, \ttt{lanl:hasFriend}, +, \ttt{lanl:jhw}, \ttt{lanl:hasFriend}, +, \ttt{lanl:norman}$)$. In the presented path, $g^p_0 = \ttt{lanl:marko}$,  $g^p_{1'} = \ttt{lanl:hasFriend}$, $g^p_{1''} = +$, $g^p_1 = \ttt{lanl:jhw}$, $g^p_{2'} = \ttt{lanl:hasFriend}$, $g^p_{2''} = +$, and $g^p_2 = \ttt{lanl:norman}$. Note that $g^p_{0'} = \emptyset$ and $g^p_{0''} = \emptyset$. The example path is diagrammed in Figure \ref{fig:path-example}.

The multi-set $q^p$ is an ordered set of vertices, edges, and directionalities that are recorded by $p$ along its path through $G^n$. The set $q^p$ maintains the same indexing schema of $'$ and $''$ as $g^p$. The main distinction between $g^p$ and $q^p$ is that $q^p$ is the returned path, \textit{not} the actual path of $p$. If $p$ reaches its destination \ttt{rwr:ExitContext} in $\Psi$ and thus vertex $j \in V$, then the set $q^p$ is one of the elements in the return set $Q$ of the path function $\rho$. Thus, for the grammar-based geodesic model,
\begin{equation*}
	Q = \bigcup_{p \in P} q^p : (q^p_0 = i \; \wedge \; q^p_{\frac{|q^p|-1}{3}} = j).
\end{equation*}
The $\frac{|q^p|-1}{3}$  is necessary to transform the length of $q^p$ into an index in $n$ time (due to the $'$ and $''$ notation convention) because the set $q^p$ includes edge labels and edge directionality as well as vertices.

\subsection{Entering $G^n$ and $\Psi$}

The initial walker $p$ starts its journey at the \ttt{rwr:EntryContext} in $\Psi$ and the vertex $i$ in $V$. Thus, $g^p_0 = i$. As in Figure \ref{fig:ont-rwr}, the \ttt{rwr:EntryContext} must be the domain of the predicate \ttt{rwr:forResource} whose range is $i$. An \ttt{rwr:EntryContext} must have no \ttt{rwr:Attributes} and must have the rule \ttt{rwr:PathCount} such that $q^p_0 = i$.

From $i \in V$ and the \ttt{rwr:EntryContext} in $\Psi$, $p$ will move to some new $k \in V$ and some new \ttt{rwr:Context} in $\Psi$. Before discussing the \ttt{rwr:Traverse} rule, it is necessary to discuss the attributes that determine the set of legal edges that can be traversed by $p$.

\subsection{The \ttt{rwr:NotEver} Attribute}

The \ttt{rwr:NotEver} attribute is useful for ensuring that path loops do not occur and thus cause the path algorithm to run indefinitely. If $p$ is trying to traverse to a new \ttt{rwr:Context} at $n+1$ and that \ttt{rwr:Context} has the \ttt{rwr:NotEver} attribute, then
\begin{align*}
\overline{X}(p)_{n+1} = \bigcup_{m \leq n} g^p_m .
\end{align*}
The set $\overline{X}(p)_{n+1}$ is the set of vertices in $V$ for which $p$ cannot legally resolve the $n+1$ \ttt{rwr:Context} to. Note that the definition of $\overline{X}(p)$ does not include edge labels or edge directionality, only vertices. This is due to the fact that the time index ($n$) of $g^p$ are not superscripted with $'$ or $''$.

\subsection{The \ttt{rwr:Is} Attribute}

\begin{figure}[h!]
	\begin{center}
		\includegraphics[width=0.4\textwidth]{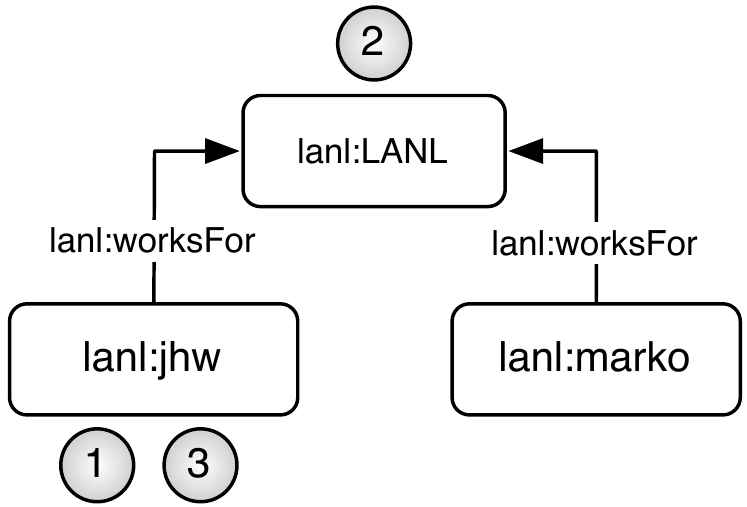}
	\caption{\label{fig:is-example}\ttt{rwr:Is} can be used to ensure that a walker backtracks.}
	\end{center}
\end{figure}

The \ttt{rwr:Is} attribute guarantees that the vertex resolved to by a particular \ttt{rwr:Context} is  a vertex seen on a previous step of the walker's $g^p$. For instance, suppose that a walker must check that a particular individual works for the Los Alamos National Laboratory before traversing a different edge label of \ttt{lanl:jhw}. This problem is diagrammed in Figure \ref{fig:is-example}.

In Figure \ref{fig:is-example}, the walker is at \ttt{lanl:jhw} at time step $n=1$. At time step $n=2$, the walker must check to see if \ttt{lanl:jhw} \ttt{lanl:worksFor} \ttt{lanl:LANL}. To do so, the walker will traverse \ttt{lanl:worksFor} edge. Upon validating the \ttt{lanl:LANL}, the walker must return back to \ttt{lanl:jhw}. Therefore, the walker will take the inverse of the \ttt{lanl:worksFor} edge (i.e.~oppose the directionality of the edge). However, despite the existence of an inverse \ttt{lanl:worksFor} edge to \ttt{lanl:marko}, the walker should not clone itself. Therefore, in order to specify that the walker must return to \ttt{lanl:jhw}, it is important to use the \ttt{rwr:Is} attribute such that only a single walker $p$ returns to \ttt{lanl:jhw} at $n=3$ and $P$ is unchanged.

The set of all legal vertices that an \ttt{rwr:Context} can resolve to is defined by the set $O$, where if $\psi$ is the \ttt{rwr:Context} at $n+1$ that maintains an \ttt{rwr:Is} attribute, then
\begin{align*}
M = & \{ ?m \; | \; \la \psi, \ttt{rwr:hasAttributes}, ?x \ra \in \Psi \\
	& \; \la ?x, \ttt{rwr:hasAttribute}, ?y \ra \in \Psi \\
	& \; \la ?y, \ttt{rdf:type}, \ttt{rwr:Is} \ra \in \Psi \\
	& \; \la ?y, \ttt{rwr:step}, ?m \ra \in \Psi \}
\end{align*}
and
\begin{equation*}
O(p)_{n+1} = \bigcup_{m \in M} g^p_{n-m}.
\end{equation*}
The set $O(p) \subseteq V$ is the set of legal vertex resources that the $n+1$ \ttt{rwr:Context} $\psi$ can resolve to and is used in the calculation of an \ttt{rwr:Traverse} at $n$.

\subsection{The \ttt{rwr:Not} Attribute}

\begin{figure}[h!]
	\begin{center}
		\includegraphics[width=0.475\textwidth]{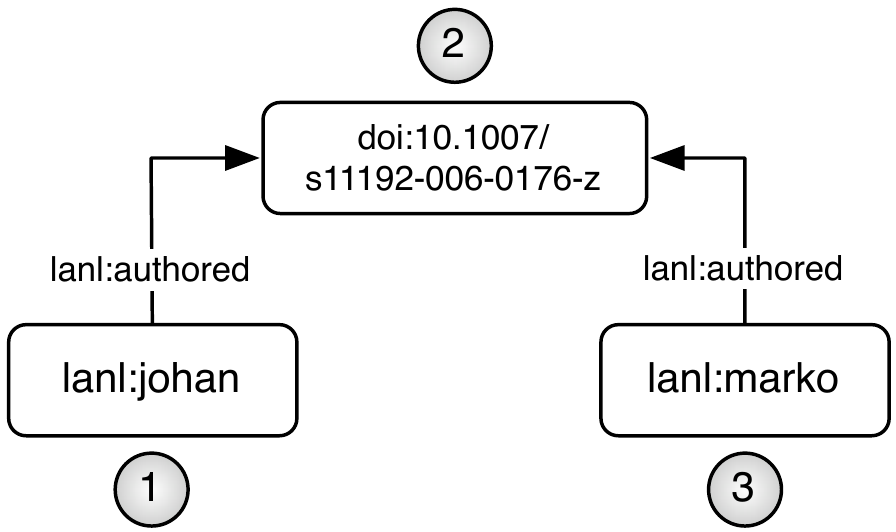}
	\caption{\label{fig:not-example}\ttt{rwr:Not} can be used to ensure that a walker does not backtrack.}
	\end{center}
\end{figure}

The \ttt{rwr:Not} attribute determines the set of vertices that the $n+1$ \ttt{rwr:Context} cannot resolve to. This is similar to the $\overline{X}(p)$ set, except that it is for some $n$, not for all $n$ in the past. For example, suppose that the walker must only consider an article co-authorship network. This problem is diagrammed in Figure \ref{fig:not-example}.

In Figure \ref{fig:not-example}, the walker must determine if the article \ttt{doi:10.1007/s11192-006-0176-z} has at least 2 co-authors. In order to do so, the walker must not return to \ttt{lanl:jbollen} at $n=3$. If
\begin{align*}
M = & \{ ?m \; | \; \la \psi, \ttt{rwr:hasAttributes}, ?x \ra \in \Psi \\
	& \; \la ?x, \ttt{rwr:hasAttribute}, ?y \ra \in \Psi \\
	& \; \la ?y, \ttt{rdf:type}, \ttt{rwr:Not} \ra \in \Psi \\
	& \; \la ?y, \ttt{rwr:step}, ?m \ra \in \Psi \}
\end{align*}
and
\begin{equation*}
X(p)_{n+1} = \bigcup_{m \in M} g^p_{n-m},
\end{equation*}
then $X(p) \subseteq V$ is the set of vertices that the $n+1$ \ttt{rwr:Context} $\psi$ must not resolve to and is used in the calculation of an \ttt{rwr:Traverse} at $n$.

\subsection{The \ttt{rwr:Traverse} Rule}

The \ttt{rwr:Traverse} rule is perhaps the most important aspect of the grammar. An \ttt{rwr:Traverse} rule of an \ttt{rwr:Context} determines the next \ttt{rwr:Context} that $p$ should traverse to in $\Psi$ as well as the next $k \in V$. It utilizes the previously defined attribute sets $\overline{X}(p)$, $O(p)$, and $X(p)$ in its calculation. An \ttt{rwr:Traverse} rule is composed of a set of \ttt{rwr:Edge}s that can be either incoming or outgoing. Thus, unlike in directed networks, the path of a $p$ is not constrained by the directionality of the edges. The $\Gamma$ functions are defined as $\Gamma : V \times P \rar G$ and $t$ is the \ttt{rwr:Traverse} rule of the current \ttt{rwr:Context} $\psi$. Therefore, if
\begin{align*}
Y_{\text{out}} = & \{ ?y \; | \; \la t, \ttt{rwr:hasEdge}, ?y \ra \in \Psi \\
			& \; \la ?y, \ttt{rdf:type}, \ttt{rwr:OutEdge} \ra \in \Psi \},
\end{align*}
\begin{align*}
Y_{\text{in}} = & \{ ?y \; | \; \la t, \ttt{rwr:hasEdge}, ?y \ra \in \Psi \\
			& \; \la ?y, \ttt{rdf:type}, \ttt{rwr:InEdge} \ra \in \Psi \},
\end{align*}
\begin{align*}
\Gamma^+(a,p) = \bigcup_{y \in Y_{\text{out}}} & \{ \la a,?\omega,?b \ra \; | \; \la a, ?\omega, ?b \ra \in G^n \\
				& \; \wedge \; \la y, \ttt{rwr:hasPredicate}, ?w \ra \in \Psi \\
				& \; \wedge \; (\la ?\omega, \ttt{rdfs:subPropertyOf}, ?w \ra \in G^n \\
				& \; \; \; \; \; \; \; \vee \; ?\omega = ?w) \\
				& \; \wedge \; \la y, \ttt{rwr:hasObject}, ?x \ra \in \Psi \\
				& \; \wedge \; \la ?x, \ttt{rwr:forResource}, ?z \ra \in \Psi \\
				& \; \wedge \; (\la ?b, \ttt{rdf:type}, ?z \ra \in G^n \; \vee \; ?b = ?z) \\
				& \; \wedge \; (O(p)_{n+1} = \emptyset \; \vee \; ?b \in O(p)_{n+1}) \\
				& \; \wedge \; ?b \notin X(p)_{n+1} \; \wedge \; ?b \notin \overline{X}(p)_{n+1} \},
\end{align*}
and
\begin{align*}
\Gamma^-(a,p) = \bigcup_{y \in Y_{\text{in}}} & \{ \la ?b,?\omega,a \ra \; | \; \la ?b, ?\omega, a \ra \in G^n \\
				& \; \wedge \; \la y, \ttt{rwr:hasPredicate}, ?w \ra \in \Psi \\
				& \; \wedge \; (\la ?\omega, \ttt{rdfs:subPropertyOf}, ?w \ra \in G^n \\
				& \; \; \; \; \; \; \; \vee \; ?\omega = ?w) \\
				& \; \wedge \; \la y, \ttt{rwr:hasSubject}, ?x \ra \in \Psi \\
				& \; \wedge \; \la ?x, \ttt{rwr:forResource}, ?z \ra \in \Psi \\
				& \; \wedge \; (\la ?b, \ttt{rdf:type}, ?z \ra \in G^n \; \vee \; ?b = ?z) \\
				& \; \wedge \; (O(p)_{n+1} = \emptyset \; \vee \; ?b \in O(p)_{n+1}) \\
				& \; \wedge \; ?b \notin X(p)_{n+1} \; \wedge \; ?b \notin \overline{X}(p)_{n+1} \},
\end{align*}
then
\begin{equation*}
	\Gamma(a,p) = \Gamma^+(a,p) \cup \Gamma^-(a,p),
\end{equation*}
where $\Gamma(a,p)$ is the set of legal edges that $p$ can traverse given its current $V$ location of $a$ and $\Psi$ location $\psi$. Note that the set $\Gamma(a,p)$ has a unique set of elements. If $\Gamma(a,p) = \emptyset$, then $p$ halts.

Unlike the grammar-based eigenvector model of \cite{grammar:rodriguez2008}, the geodesic requires the searching of all legal paths. In line with a breadth-first search, all network branches are checked. Thus, for every triple $\tau \in \Gamma(a,p)$, a clone walker is created and added to $P$. This idea will be made more salient in the example to follow.

\subsection{The \ttt{rwr:PathCount} Rule}

The \ttt{rwr:PathCount} rule is the mechanism by which values in $g^p$ get appended to $q^p$, where $q^p$ is the path returned by $p$ at the end of the algorithm's execution. The rule instructs $p$ to append a path segment in $g^p$ to the ordered multi-set $q^p$. If a particular \ttt{rwr:Context} $\psi$ has the \ttt{rwr:PathCount} rule with the \ttt{rwr:step} $x$ such that $x \in \mb{N}$, then $p$ will append $g^p_{n-x'}$, $g^p_{n-x''}$, and $g^p_{n-x}$ to $q^p$ such that none of the elements copied from $g^p = \emptyset$ and they are added in their respective order. 

The next section will present the aforementioned rules and attributes within the framework of a particular social network ontology in order to demonstrate a practical application.

\section{Geodesics in a Semantic Social Network}

\begin{figure}[h!]
	\begin{center}
		\includegraphics[width=0.5\textwidth]{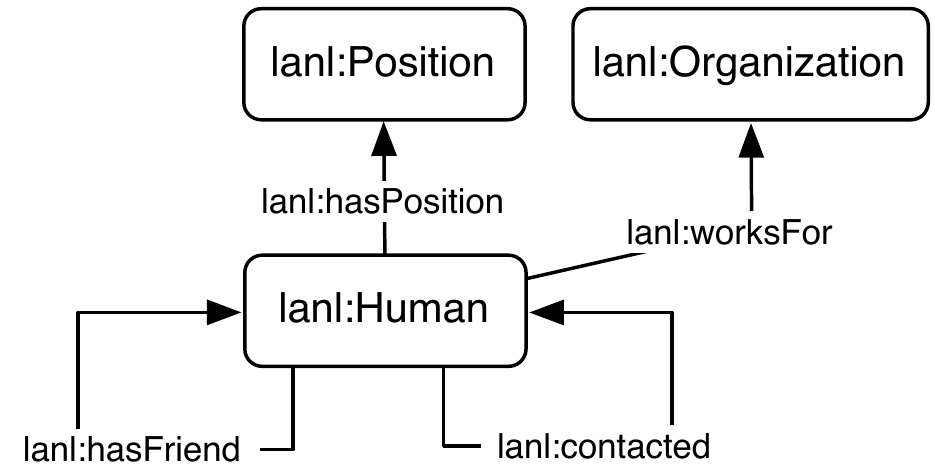}
	\caption{\label{fig:social-ontology}An example semantic social network ontology.}
	\end{center}
\end{figure}

This section will present two examples of the previously presented ideas to the problem of calculating semantically meaningful geodesic functions within a semantic social network. Figure \ref{fig:social-ontology} presents an RDFS network ontology that will be used throughout the remainder of this section. Note that the domain and range of the properties are denoted by the tail and head of the edge, respectively.

\begin{figure}[h!]
	\begin{center}
		\includegraphics[width=0.95\textwidth]{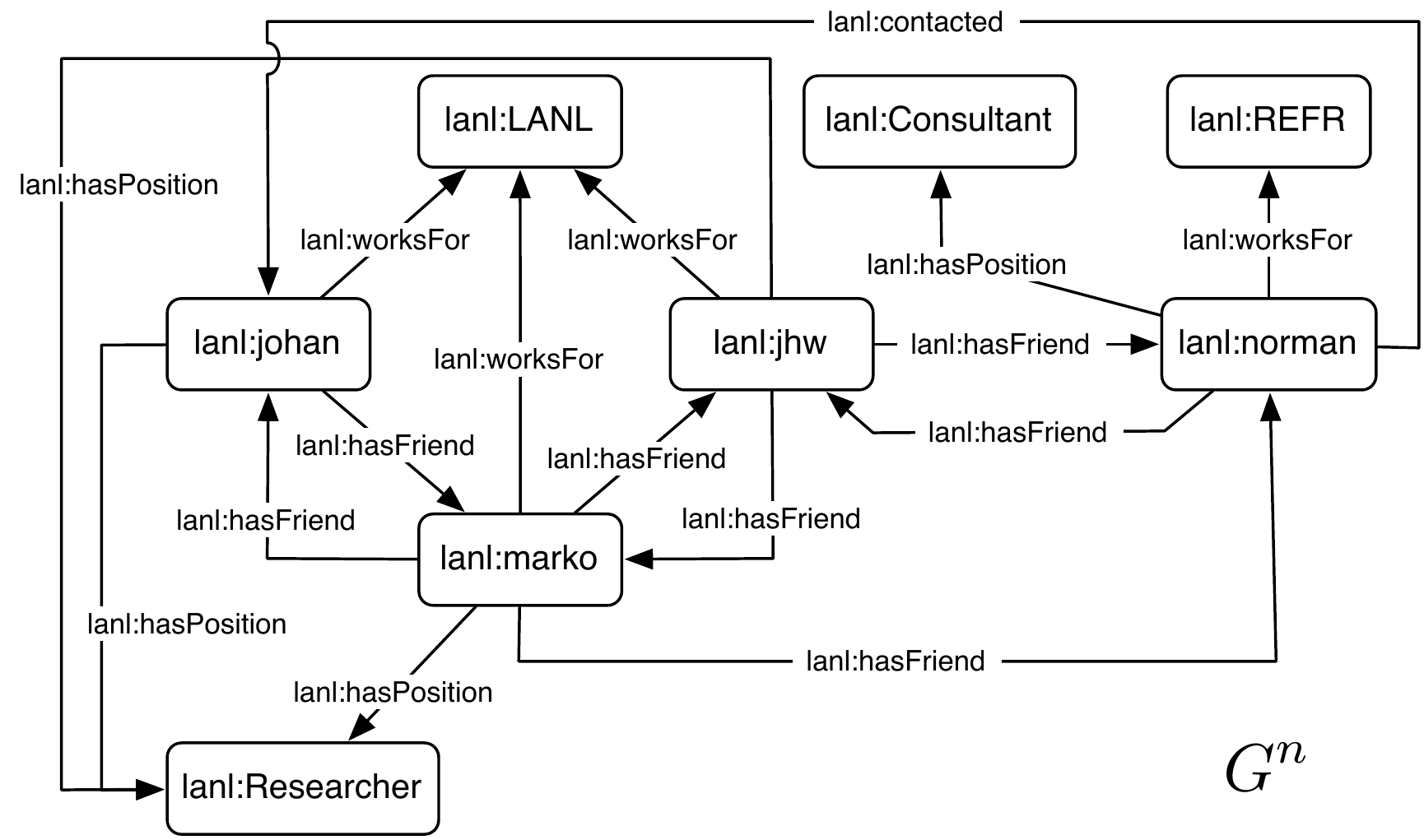}
	\caption{\label{fig:social-instance}An example semantic social network instance.}
	\end{center}
\end{figure}

Figure \ref{fig:social-instance} diagrams an example instance that respects the ontological constraints diagrammed in Figure \ref{fig:social-ontology}.

The first example will demonstrate how to determine all the non-recurrent paths between the vertex \ttt{lanl:johan} and \ttt{lanl:norman} such that only friendship paths are taken, but those intervening friend vertices must have a \ttt{lanl:Researcher} position. The second example will present a grammar that simulates an unlabeled network path calculation by ignoring vertex types and edge labels. 

Note that the two examples presented are for locating all paths between a source and a sink vertex. This is for demonstration purposes only. If one required only the shortest path, once a path between the source and sink has been found, the algorithm can halt. In unweighted networks, using a breadth-first search algorithm, the first path discovered is always the shortest path \cite{algs:cormen1999}.

\subsection{A Non-Recurrent Paths Grammar\label{sec:first-example}}

\begin{figure}[h!]
	\begin{center}
		\includegraphics[width=0.8\textwidth]{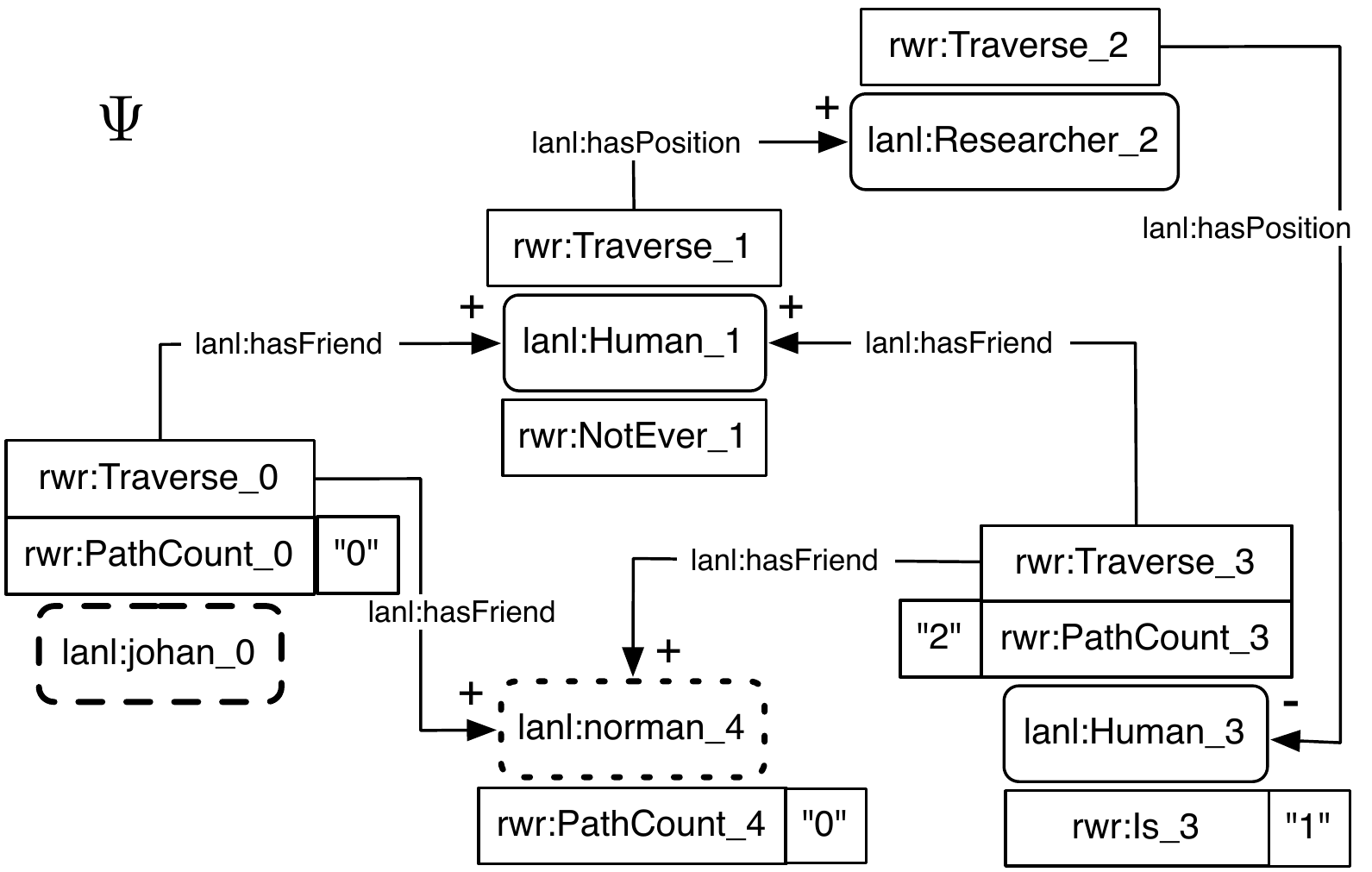}
	\caption{\label{fig:path1-example}A grammar to determine all non-recurrent \ttt{lanl:hasFriend} paths from \ttt{lanl:johan} to \ttt{lanl:norman}.}
	\end{center}
\end{figure}

Figure \ref{fig:path1-example} presents a geodesic grammar that determines the set of all non-recurrent paths between \ttt{lanl:johan} and \ttt{lanl:norman} according to \ttt{lanl:hasFriend} relationships where every friend along the walker's path must be a \ttt{lanl:Researcher}. 

Note the diagrammatic conventions used to represent a grammar. Every \ttt{rwr:Context}, \ttt{rwr:Rule}, and \ttt{rwr:Attribute} has a \_\# after its type. This is to denote that each representation of the same \ttt{rwr:Context}, \ttt{rwr:Rule}, or \ttt{rwr:Attribute} is, in fact, a distinct vertex in $\Psi$. The label of the \ttt{rwr:Context} is the object of the \ttt{rwr:forResource} property minus the \_\#. Furthermore, the dashed contexts are \ttt{rwr:EntryContext}s and the dotted contexts are \ttt{rwr:ExitContext}s. Thus, \ttt{lanl:johan\_0} is the source context and \ttt{lanl:norman\_4} is the sink context in $\Psi$, and where \ttt{lanl:johan} is the source vertex and \ttt{lanl:norman} is the sink vertex in $G^n$.

The \ttt{rwr:Rules} of an \ttt{rwr:Context} are represented in their order of execution from bottom to top. The \ttt{rwr:Attributes} are associated, in no particular order, with their respective \ttt{rwr:Context}. If a rule or attribute requires a literal \ttt{rwr:step} specification, that literal is appended to its respective rule or attribute. The \ttt{+} or \ttt{-} symbol on the head of an edge denotes whether the \ttt{rwr:Traverse} edge is an \ttt{rwr:OutEdge} or \ttt{rwr:InEdge}, respectively.

At $n=0$, $g^{p_0}_0 = \ttt{lanl:johan}$ and $P = \{ p_0 \}$.  The first rule to be executed is the \ttt{rwr:PathCount\_0} rule in which $p_0$ will register $g^{p_0}_0$ in $q^p$ such that  $q^{p_0}_0 = g^{p_0}_0$. After adding \ttt{lanl:johan} to $q^{p_0}$, the walker will execute the \ttt{rwr:Traverse\_0} rule. The \ttt{rwr:Traverse\_0} rule yields a $\Gamma(\ttt{lanl:johan}, p_0) = \{ \la \ttt{lanl:johan}, \ttt{lanl:hasFriend}, \ttt{lanl:marko} \ra \}$. If \ttt{lanl:norman} was a friend of \ttt{lanl:johan}, then that edge would have been represented in $\Gamma(\ttt{lanl:johan},p_0)$ as well. Because $\ttt{lanl:marko} \notin g^{p_0}$, the \ttt{rwr:NotEver\_1} attribute of the \ttt{Human\_1} context has an $\overline{X}(p_0)_1 = \emptyset$.

At $n=1$, the current path of $p_0$ is $g^{p_0} = (\ttt{lanl:johan}, \ttt{lanl:hasFriend}, +,  \ttt{lanl:marko})$ and the current return path $q^{p_0} = (\ttt{lanl:johan})$. There exists only one rule at \ttt{rwr:Human\_1}. The \ttt{rwr:Traverse\_1} rule dictates that $p_0$ take an outgoing edge from \ttt{lanl:marko} to a \ttt{lanl:Researcher} position. Given that there is only one edge that can be traversed, $\Gamma(\ttt{lanl:marko},p_0) = \{ \la \ttt{lanl:marko}, \ttt{lanl:hasPosition}, \ttt{lanl:Researcher} \ra \}$.

At $n=2$, the current path of $p_0$ is $g^{p_0} = ($\ttt{lanl:johan}, \ttt{lanl:hasFriend}, +,  \ttt{lanl:marko}, \ttt{lanl:hasPosition}, +, \ttt{lanl:Researcher}$)$ and the current return path $q^{p_0} = (\ttt{lanl:johan})$. The only rule of the \ttt{lanl:Researcher\_2} context is to return the human that was last encountered as specified by the \ttt{rwr:Is\_3} attribute of the next \ttt{lanl:Human\_3} context. Thus, $\Gamma(\ttt{lanl:Researcher},p_0) = \{ \la \ttt{lanl:marko}, \ttt{lanl:hasPosition}, \ttt{lanl:Researcher} \ra \}$.

At $n=3$, the current path of $p_0$ is $g^{p_0} = ($\ttt{lanl:johan}, \ttt{lanl:hasFriend}, +,  \ttt{lanl:marko}, \ttt{lanl:hasPosition}, +, \ttt{lanl:Researcher}, \ttt{lanl:hasPosition}, --, \ttt{lanl:marko}$)$. Given the \ttt{rwr:PathCount\_3} rule with a \ttt{rwr:step} of $2$, $q^{p_0} = ($\ttt{lanl:johan}, \ttt{lanl:hasFriend}, +, \ttt{lanl:marko}$)$. The \ttt{rwr:Traverse\_3} rule provides a $\Gamma(\ttt{lanl:marko}, p_0)$ with two edges such that $\Gamma(\ttt{lanl:marko},p_0) = ( \la$\ttt{lanl:marko}, \ttt{lanl:hasFriend}, \ttt{lanl:jhw}$\ra, \la$\ttt{lanl:marko}, \ttt{lanl:hasFriend}, \ttt{lanl:norman}$\ra)$. Note that the edge $\la$\ttt{lanl:marko}, \ttt{lanl:hasFriend}, \ttt{lanl:johan}$\ra$ does not exist in $\Gamma(\ttt{lanl:marko},p_0)$ because of the \ttt{rwr:NotEver\_1} attribute at the \ttt{lanl:Human\_1} context (i.e.~$\overline{X}(p_0)_4=\{ \ttt{lanl:johan},\ttt{lanl:marko} \}$). Because two edges exist in $\Gamma(\ttt{lanl:marko},p_0)$, $p_0$ is cloned such that $P=\{p_0, p_1\}$, $g^{p_0} = g^{p_1}$, and $q^{p_0} = q^{p_1}$. The walker $p_0$ will take one edge and $p_1$ will take the other edge.

At $n=4$, $p_1$ will be at \ttt{lanl:norman} in $G^n$ and thus at an \ttt{rwr:ExitContext} in $\Psi$. However, before $p_1$ halts, \ttt{rwr:PathCount\_4} is executed such that $Q = \{q^{p_1}\} = \{($\ttt{lanl:johan}, \ttt{lanl:hasFriend}, +, \ttt{lanl:marko}, \ttt{hasFriend}, +, \ttt{lanl:norman}$)\}$. At the completion of \ttt{rwr:PathCount\_4} there are no other rules to execute and thus $p_1$ halts. The walker $p_0$, on the other hand, will be at \ttt{lanl:jhw} at $n=4$. It is not until $n=7$ that $p_0$ arrives at \ttt{lanl:norman}. 

At $n=7$, $q^{p_0} = ($\ttt{lanl:johan}, \ttt{lanl:hasFriend}, +, \ttt{lanl:marko}, \ttt{lanl:hasFriend}, +, \ttt{lanl:jwh}, \ttt{lanl:hasFriend}, +, \ttt{lanl:norman}$)$. At $n=7$, the grammar is complete and $|Q| = 2$.

The shortest path of $Q$ is defined as the function $s: V \times V \times \Psi \rar \mb{N}$, where
\begin{equation*}
	s(i,j,\Psi) = min\left( \bigcup_{q \in \rho(i,j,\Psi)} \frac{|q|-1}{3} \right).
\end{equation*}
The $1$ must be subtracted from $|q|$ in order to not include source vertex $i$ as a step and then must be divided by $3$ so as to avoid the inclusion of the edge label and directionality of the edge in the path length calculation. In the example presented, the shortest ``researcher-constrained friendship" path is $2$. From $s$, it is possible to generate all other geodesic functions as defined in Section \ref{sec:geos}.

In the presented example, the source vertex is \ttt{lanl:johan} and the sink vertex is \ttt{lanl:norman}. It is noted that the \ttt{rwr:EntryContext} and \ttt{rwr:ExitContext} of $\Psi$ can be reconfigured to support new $i$ and $j$ source and sink vertices. In other words, $\Psi$ can be configured to support different $i$/$j$ path calculations.

\subsection{A Grammar to Simulate Unlabeled Geodesics}

\begin{figure}[h!]
	\begin{center}
		\includegraphics[width=0.7\textwidth]{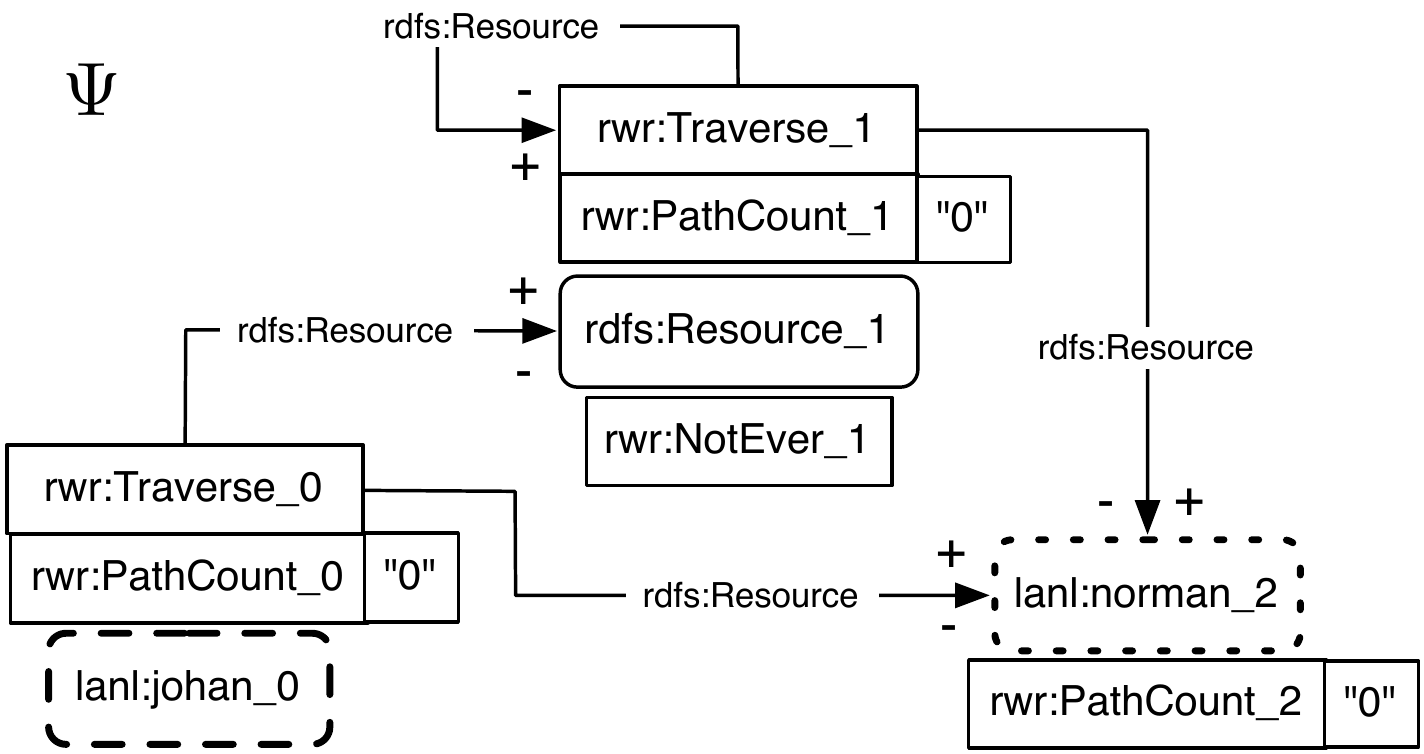}
	\caption{\label{fig:path2-example}An unconstrained grammar to determine all non-recurrent paths from \ttt{lanl:jbollen} to \ttt{lanl:norman}.}
	\end{center}
\end{figure}

This section presents another example of the grammar-based geodesic algorithm. In this example, the grammar presented is equivalent to removing the edge labels and directionality from the semantic network and calculating a traditional geodesic metric on it. Figure \ref{fig:path2-example} presents the grammar where, in RDFS, \ttt{rdfs:Resource} is the base type of all resources (vertices and edge labels). Thus, all \ttt{rwr:Context}s and \ttt{rwr:Edge}s can legally resolve to any vertex and edge label, respectively.

The grammar in Figure \ref{fig:path2-example} will determine the set of all non-recurrent paths between \ttt{lanl:johan} and \ttt{lanl:norman} such that any edge type can be traversed to any vertex type. The central \ttt{rwr:Context} is the \ttt{rdfs:Resource\_1} context. A walker will loop over \ttt{rwr:Resource\_1} until it can find an edge to make the final traversal to \ttt{lanl:norman}. Note the use of both \ttt{rwr:OutEdge}s (\ttt{+}) and \ttt{rwr:InEdge}s (\ttt{-}). With both edges accessible, the walker can walk in any direction on the network. Thus, this grammar is equivalent to executing a geodesic on an undirected and unlabeled version of the semantic network. Finally, the grammar will produce no recurrent paths because of the \ttt{rwr:NotEver\_1} rule.

Given this $\Psi$ and the original social network instance $G^n$ diagrammed in Figure \ref{fig:social-instance}, the shortest path between \ttt{lanl:johan} and \ttt{lanl:norman} is $($\ttt{lanl:johan}, \ttt{lanl:contacted}, --, \ttt{lanl:norman}$)$ with a path length of $1$. To contrast, in the first example when the walker's path was constrained to researcher friendship relationships, the shortest path between \ttt{lanl:johan} and \ttt{lanl:norman} was $2$.

\section{Analysis}

The semantic network is an unweighted network. Thus, determining the shortest path between any two vertices is best solved by a breadth-first algorithm. The grammar-based walker, through cloning, is analogous to a breadth-first search through the network. However, not all edges are considered by the walker and thus, the running time of the algorithm is less than or equal to $O(|V| + |G^n|)$. The determination of the running time of the algorithm is grammar dependent. In order to calculate the running time of a particular grammar, it is important to calculate the number of vertices and edges of the grammar-specified types in $G^n$. In the worst case situation, the walker population $P$ will have traversed all vertices and edges from the source to ultimately locate the sink. However, because the network is unweighted, once the sink has been found by a single $p \in P$, the shortest path has been determined so the algorithm is complete.

\section{Computational Reuse with $p$-Encodings}

Once a computation has been performed, its results can be reused as a sub-solution to a larger problem. As stated previously, the path calculations between two vertices in a network are the kernel calculations for more complex path metrics such as shortest path, eccentricity, radius, diameter, closeness centrality, and betweenness centrality. This section will demonstrate how to encode the $q^p$ data structure into a semantic network such that the results of these calculations can be reused for each of the higher-order metrics.

For instance, suppose the function $f : \mb{N} \rar \mb{N}$, where $f(n) = n + 1$. Furthermore, suppose that there exist the resources \ttt{"1"$^\wedge$$^\wedge$xsd:int} and \ttt{"2"$^\wedge$$^\wedge$xsd:int} such that there also exists the triple \footnote{The namespace prefix \ttt{xsd} is used to specify the data type of the quoted symbols. In this case, \ttt{xsd:int} refers to an integer data type.} 
\begin{equation*}
\la \ttt{"1"$^\wedge$$^\wedge$xsd:int}, f, \ttt{"2"$^\wedge$$^\wedge$xsd:int} \ra.
\end{equation*}
The triple  states, in human language, that the number $1$ is related to the number $2$ by the functional relationship $f$. If that triple is in $G^n$, then never again would it be necessary to compute $f(1)$ because the result has already been computed and has been represented in $G^n$. Thus, $G^n$ can be queried for the result of the $f(1)$ computation. For example,
\begin{equation*}
X = \{ ?x1 \; | \; \la \ttt{"1"$^\wedge$$^\wedge$xsd:int}, f, ?x1 \ra \in G^n \}
\end{equation*}
would return the result of $f(1)$. However, this is a trivial example because it is faster to compute $f(1)$ on the local hardware processor then it is to query $G^n$ for the solution. In other situations, this is not necessarily the case.

\begin{figure}[h!]
	\begin{center}
		\includegraphics[width=0.6\textwidth]{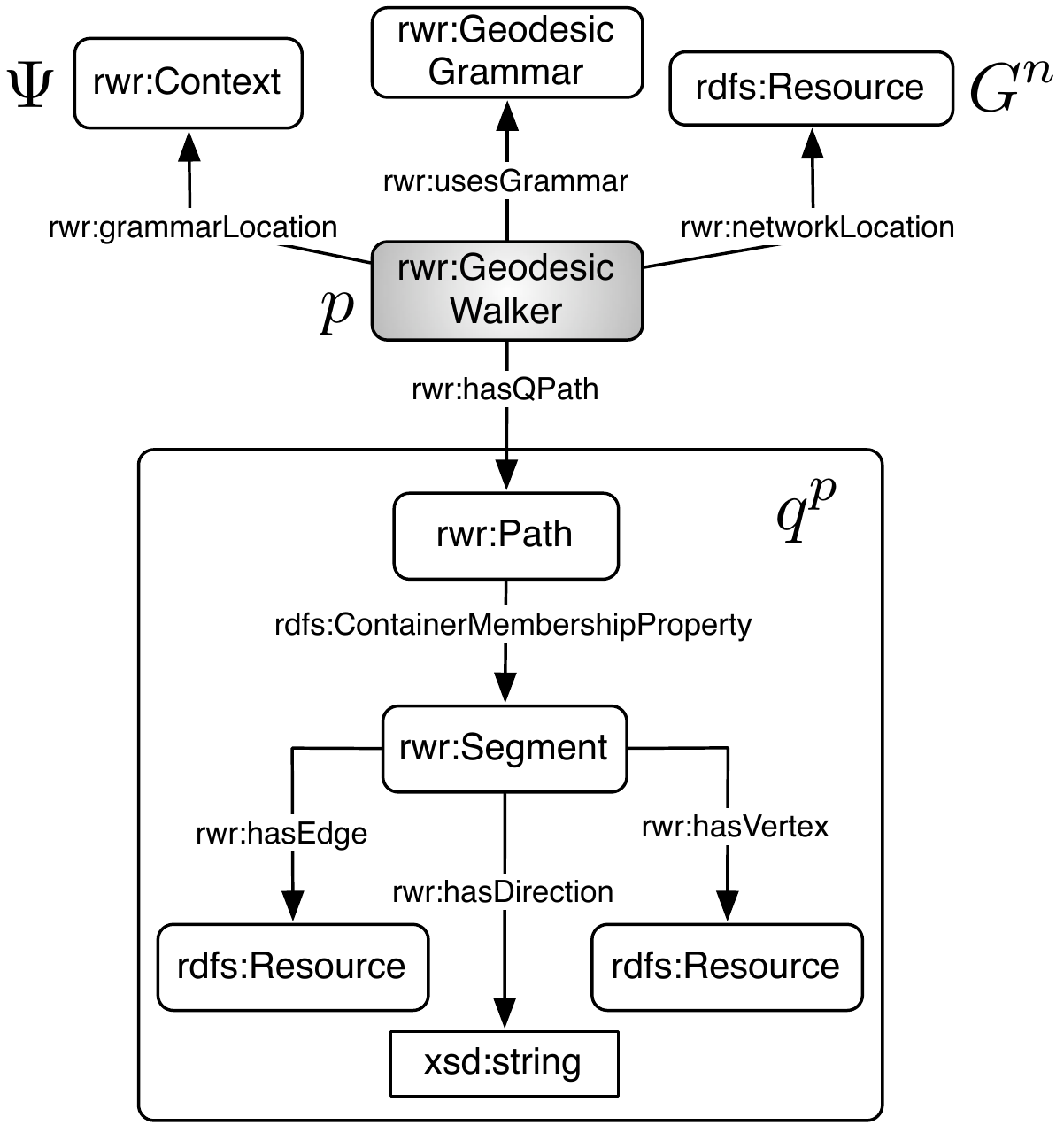}
	\caption{\label{fig:georw}Encoding $p$ and its associated $q^p$ data structure in a semantic network.}
	\end{center}
\end{figure}

For more complex computations, such as the set of paths between two vertices in $V$ according to some $\Psi$, it is possible to represent $p$ and its associated data structure $q^p$ as a semantic network. Figure \ref{fig:georw} is a diagram of the RDFS ontology representing $p$ and $q^p$, where the noted components are considered either named graphs \cite{named:carroll2005}, separate semantic network instances, or reified sub-networks \cite{rdfspec:manola2004}. From instances of this ontology, it is possible to reuse the path calculations to determine various geodesics without recalculating the $\Psi$-correct paths between any two vertices $i$ and $j$.

\begin{figure}[h!]
	\begin{center}
		\includegraphics[width=0.95\textwidth]{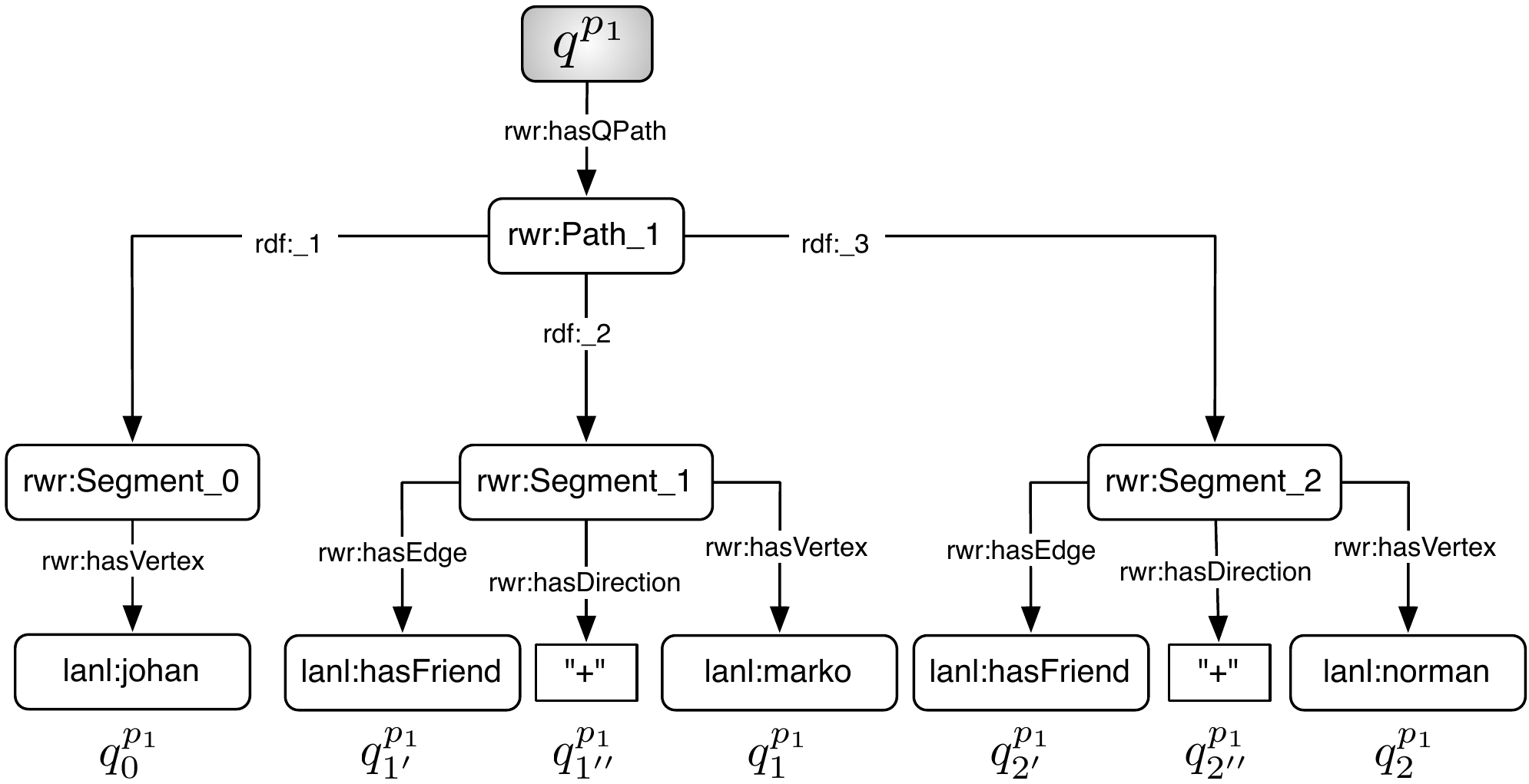}
	\caption{\label{fig:georw-example}An instance of the RDFS ontology in Figure \ref{fig:georw}.}
	\end{center}
\end{figure}

For example, given the $q^{p_1}$ path calculated in Section \ref{sec:first-example}, the semantic network representation would be represented as diagrammed in Figure \ref{fig:georw-example}. The number of \ttt{rwr:Segments} is the largest \ttt{rdfs:ContainerMembershipProperty} (i.e.~\ttt{rdf:\_3}) for the \ttt{rwr:Path}. The path length of $q^{p_1}$ is thus, $\ttt{rdf:\_3} - 1$ (i.e.~$3-1$). To make the mapping to the convention used in Section \ref{sec:first-example} more salient, note the \ttt{rwr:Segment} component labels at the bottom of the diagram.

If the grammar-based path algorithm halts when it reaches an \ttt{rwr:ExitContext}, then every $q^p$ instance is a shortest path. While only the shortest path between two vertices is required for geodesic metrics, the next subsections present the generalized algorithm for searching all $q^p$ paths between source vertex $i$ and sink vertex $j$.

\subsection{$p$-Encoded Shortest Path}

To compute the shortest path between two vertices $i$ and $j$, where the complete set $P$ is searched, the grammar-based shortest path algorithm is represented as
\begin{equation*}
s(i,j,\Psi) = min(X_{i,j}) - 1,
\end{equation*}
where
\begin{align*}
X_{i,j} =& \{ ?x2, ?x5 \; | \; \la ?x1, \ttt{rdf:type}, \ttt{rwr:GeodesicWalker} \ra \in p \\
      & \; \wedge \; \la ?x1, \ttt{rwr:usesGrammar}, \Psi \ra \in p \\
      & \; \wedge \; \la ?x1, \ttt{rwr:hasQPath}, ?x2 \ra \in p \\
      & \; \wedge \; \la ?x2, \ttt{rdf:\_1}, ?x3 \ra \in q^p \\
      & \; \wedge \; \la ?x3, \ttt{rwr:hasVertex}, i \ra \in q^p \\
      & \; \wedge \; \la ?x4, \ttt{rdf:type}, \ttt{rwr:EntryContext} \ra \in \Psi \\
      & \; \wedge \; \la ?x4, \ttt{rwr:forResource}, i \ra \in \Psi \\
      & \; \wedge \; \la ?x2, ?x5, ?x6 \ra  \in q^p \\
      & \; \wedge \; \la ?x6, \ttt{rwr:hasVertex}, j \ra  \in q^p \\
      & \; \wedge \; \la ?x7, \ttt{rdf:type}, \ttt{rwr:ExitContext} \ra \in \Psi \\
      & \; \wedge \; \la ?x7, \ttt{rwr:forResource}, j \ra \in \Psi \}
\end{align*}
and the function $min: \ttt{rwr:Path} \times \ttt{rdfs:ContainerMembershipProperty} \rar \mb{N}$ returns the smallest value of the second component of its domain minus the \ttt{rdf:\_} head. For example, if $X_{i,j} = \{(\ttt{rwr:Path\_0},\ttt{rdf:\_4}),(\ttt{rwr:Path\_1}, \ttt{rdf:\_3})\}$, then $min(X_{i,j}) = 3$. The first \ttt{rwr:Path} element is used later when calculating the betweenness centrality of a vertex.

The $X_{i,j}$ query simply returns the path identifier and the number of segments of each path between the \ttt{rwr:EntryContext} and the \ttt{rwr:ExitContext}. More specifically, the query that generates $X_{i,j}$ can be understood, in human language, as saying: ``Given the set of all \ttt{rwr:GeodesicWalker}s ($?x1$) that use $\Psi$ as their grammar and who have a $q$-path ($?x2$) that has $i$ as the vertex of the first (i.e.~\ttt{rdf:\_1}) \ttt{rwr:Segment} ($?x3$), where $i$ is the \ttt{rwr:EntryContext} vertex of $\Psi$ ($?x4$) and who have $j$ in a $q$-path \ttt{rwr:Segment} ($?x6$), where $j$ is the \ttt{rwr:ExitContext} vertex of $\Psi$ ($?x7$), return the \ttt{rwr:Path} ($?x2$) and the \ttt{rwr:Segment} count ($?x5$) of the $j$ \ttt{rwr:Segment}."

\subsection{$p$-Encoded Eccentricity, Radius, and Diameter}

Given the shortest path query, it is possible to generate other grammar-based geodesics. For instance, for eccentricity,
\begin{equation*}
e(i,\Psi) = max\left(\bigcup_{j \in V} s(i,j,\Psi)\right).
\end{equation*}

For radius,
\begin{equation*}
r(G^n,\Psi) = min\left(\bigcup_{i \in V} e(i,\Psi)\right).
\end{equation*}

Finally, for diameter,
\begin{equation*}
d(G^n,\Psi) = max\left(\bigcup_{i \in V} e(i,\Psi)\right).
\end{equation*}

\subsection{$p$-Encoded Closeness and Betweenness Centrality}

For closeness centrality,
\begin{equation*}
	c(i,\Psi) = \frac{1}{\sum_{j \in V} s(i,j,\Psi)}.
\end{equation*}

Finally, for betweenness centrality, if $ms : \ttt{rwr:Path} \times \ttt{rdfs:ContainerMembershipProperty} \rar \ttt{rwr:Path}$, where $ms$ returns the set of shortest paths in its domain and
\begin{align*}
Y_{j,k,i} =& \{ ?x2 \; | \; \la ?x1, \ttt{rdf:type}, \ttt{rwr:GeodesicWalker} \ra \in p \\
      & \; \wedge \; \la ?x1, \ttt{rwr:usesGrammar}, \Psi \ra \in p \\
      & \; \wedge \; \la ?x1, \ttt{rwr:hasQPath}, ?x2 \ra \in p \\
      & \; \wedge \; \la ?x2, \ttt{rdf:\_1}, ?x3 \ra \in q^p \\
      & \; \wedge \; \la ?x3, \ttt{rwr:hasVertex}, j \ra \in q^p \\
      & \; \wedge \; \la ?x4, \ttt{rdf:type}, \ttt{rwr:EntryContext} \ra \in \Psi \\
      & \; \wedge \; \la ?x4, \ttt{rwr:forResource}, j \ra \in \Psi \\
      & \; \wedge \; \la ?x2, ?x5, ?x6 \ra  \in q^p \\
      & \; \wedge \; \la ?x6, \ttt{rwr:hasVertex}, i \ra  \in q^p \\
      & \; \wedge \; \la ?x2, ?x7, ?x8 \ra  \in q^p \\
      & \; \wedge \; \la ?x8, \ttt{rwr:hasVertex}, k \ra  \in q^p \\      
      & \; \wedge \; \la ?x9, \ttt{rdf:type}, \ttt{rwr:ExitContext} \ra \in \Psi \\
      & \; \wedge \; \la ?x9, \ttt{rwr:forResource}, k \ra \in \Psi \\
      & \; \wedge \; \; ?x7 \; > \; ?x5 \\
      & \; \wedge \; ?x2 \in ms(X_{j,k}) \}
\end{align*}
represents the set of shortest paths from $j$ to $k$ such that there exists some \ttt{rwr:Segment} in the \ttt{rwr:Path} that has $i$ as its vertex, then
\begin{equation*}
	b(i,\Psi) = \sum_{i \neq j \neq k \in V} \frac{|Y_{j,k,i}|}{|ms(X_{j,k})|}.
\end{equation*}

To calculate the betweenness centrality of vertex $i$, it is important to know the number of shortest paths that go from $j$ to $k$ as well as the number of shortest paths that go from $j$ to $k$ through $i$. The function $ms$  is used to determine which of those elements in $X_{j,k}$ are shortest paths. The set $Y_{j,k,i}$ is then the set of all paths between $j$ and $k$ that go through $i$ and are elements of $ms(X_{j,k})$.

\section{Conclusion}

This article has presented a technique to port some of the most fundamental geodesic network analysis algorithms into the semantic network domain. There currently exist many technologies to support large-scale semantic network models represented according to RDF. High-end, modern-day triple-stores support on the order of $10^9$ triples  \cite{lee:triple2004,hexastore:weiss2008}. While many centrality algorithms are costly on large networks, by restricting the search to meaningful subsets of the full semantic network, as defined by a grammar, geodesic metrics can be reasonably executed on even the most immense and complex of data sets \cite{semever:bollen2007,bollen:mesur2008}.

\section*{Acknowledgments}

Marko A. Rodriguez is funded by the MESUR project (http://www.mesur.org) which is supported  by a grant from the Andrew W. Mellon Foundation. Marko is also funded by a Director's Fellowship granted by the Los Alamos National Laboratory.

\end{document}